# Single Photon Generation in a Cavity Optomechanical System induced by a Nonlinear Photonic Crystal

Bijita Sarma and Amarendra K. Sarma*

*Abstract*— We propose a novel scheme for realizing single-photon blockade in a weakly driven hybrid cavity optomechanical system consisting of a nonlinear photonic crystal. Sub-Poissonian statistics is realized even when the single-photon optomechanical coupling strength is smaller than the decay rate of the optical mode. The scheme relaxes the requirement of strong coupling for photon blockade in optomechanical systems. It is shown that photon blockade could be generated at the telecommunication wavelength.

## I. INTRODUCTION

Generation of single photons is a pivotal requirement in quantum information and communication technologies [1,2]. The fingerprint of a single photon source is the observation of sub-Poissonian radiation when driven by a classical light field. Photon blockade is a possible mechanism, in which a strong optical nonlinearity makes the presence of one photon in the system enough to prohibit simultaneous presence of another photon. In recent years, tremendous progress has been made in demonstrating single photon sources in diverse platforms, such as diamond nanophotonic circuits [3], solid-state systems [4,5], optical fibers [6], Bose-Einstein condensate [7] and photonic crystals [8]. In most cases, the prototype system that has been widely explored is cavity quantum electrodynamics.

Recently, engineered quantum devices based on optomechanical systems have garnered much attention owing to numerous possible applications towards ground state cooling of mesoscopic oscillators, entanglement of optical and mechanical modes, non-classical state generation, quantum state transfer between different modes etc. [9,10]. A typical optomechanical system consists of an optical cavity in which one of the end-mirrors is movable by the force exerted by light. The system is intrinsically nonlinear due to the coupling between optical field and mechanical motion and exhibits Kerr-type nonlinearity. This feature of optomechanical systems has been utilized to propose single photon source or emitter by using the phenomenon of photon blockade [10-14]. However, as the effect of one photon tends to be very tiny, conventional photon blockade in optomechanical systems requires the system to be in the strong coupling regime, i.e. $g > \kappa$, where $g$ is the single-photon optomechanical coupling strength and $\kappa$ is the decay rate of the optical mode [11,12]. This strong coupling condition is hard to achieve in experiments.

In this letter, we propose to use a photonic crystal cavity showing third order nonlinearity in-between the mirrors of the cavity to circumvent the issue of weak nonlinearity. Optomechanical photonic crystals offer the privilege of added material-induced nonlinearity such as third-order nonlinear susceptibility, in addition to the inherent optomechanical Kerr-type nonlinearity [15]. Photonic crystals offer the advantage of availability of diffraction-limited mode volumes and ultrahigh quality factors. In this letter, we show that the stringent requirement to have $g > \kappa$, in order to achieve single photon blockade in an optomechanical system, could be relaxed using our proposal. Also, it is worthy to be noted that the prospective long distance communication of single-photon requires the workability around 1550 nm, i.e. the so-called telecommunication wavelength, as the photons generated should also have a wavelength that would give low loss and attenuation while travelling through an optical fiber. This issue has not been explored much in optomechanical systems. In this letter, we have considered the wavelength of the cavity radiation to be 1550 nm and show that it is possible to obtain strong photon antibunching at this wavelength.

## II. SYSTEM AND METHOD

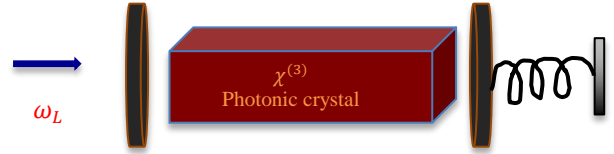

Fig. 1. Schematic diagram of a nonlinear optomechanical photonic crystal-cavity system driven by a resonating coherent field.

The schematic of the proposed system is depicted in Fig. 1, where a Kerr-type photonic crystal is inserted in an optomechanical cavity with one mirror fixed while the other one is movable. The cavity system is weakly driven by a laser with frequency $\omega_L$. The cavity field frequency is $\omega_a$. The electromagnetic field couples with the mechanical mirror motion via radiation pressure and results in an optomechanical coupling term. In a rotating frame at the laser frequency, the Hamiltonian of the driven system is given by [8,9]:

$$H = \hbar\Delta_a a^\dagger a + \hbar\omega_b b^\dagger b + U a^\dagger a^\dagger a a + \hbar g a^\dagger a(b + b^\dagger) + \hbar\Omega(a + a^\dagger). \quad (1)$$

Here, $a$ $(a^\dagger)$ and $b$ $(b^\dagger)$ are the annihilation (creation) operators of the cavity field and mechanical resonator respectively. $\Delta_a = \omega_a - \omega_L$ is the detuning of the cavity field and $\omega_b$ is the mechanical oscillator resonance frequency. $g =$

The authors are with Department of Physics, Indian Institute of Technology Guwahati, Guwahati-781039, Assam, India. * aksarma@iitg.ernet.in

$\frac{\omega_a}{L} x_{ZPF}$ is the single-photon optomechanical coupling strength, where $L$ is the length of the cavity and $x_{ZPF}$ is the zero-point fluctuation of the mechanical oscillator. $\Omega = \sqrt{\kappa P_{in}/\hbar \omega_L}$ is the drive amplitude, $P_{in}$ and $\kappa = \omega_a/Q_a$ being the input power and optical mode decay rate respectively. $Q_a$ is the quality factor of the optical cavity. $U = 3(\hbar \omega_a)^2 D/(4\varepsilon_0 V_{eff})$ is the strength of nonlinear Kerr interaction, where $D = \bar{\chi}^{(3)}/\bar{\varepsilon}_r^2$, where $\bar{\chi}^{(3)}$ and $\bar{\varepsilon}_r$ are, respectively, the average nonlinear susceptibility and relative dielectric permittivity of the photonic crystal. $V_{eff} \sim (\lambda_a/2\bar{n}_r)^3$ is the cavity mode volume, considering difraction-limited confinement volume for a Si-based system, where $\lambda_a$ is the wavelength of the cavity radiation and $\bar{n}_r = \sqrt{\bar{\varepsilon}_r}$ [16]. The first and second term in Eq. (1) gives the energy of the cavity mode and the mechanical mode respectively. The third term accounts for the interaction energy between the cavity mode and the photonic crystal. The optomechanical interaction between the cavity mode and the mechanical oscillator is described by the fourth term. The last term describes the laser driving.

The statistical properties of the photon field could be described by the normalized zero-time delay second order correlation function:

$$g^{(2)}(0) = \frac{\langle a^\dagger(t) a^\dagger(t) a(t) a(t) \rangle}{\langle a^\dagger(t) a(t) \rangle^2} \quad (2)$$

In the next section, we find out the second-order photon correlation function by numerically solving the quantum master equation of the system:

$$\frac{d\rho}{dt} = \frac{i}{\hbar}[\rho, H] + L_a(\rho) + L_b(\rho), \quad (3)$$

where $L_a(\rho) = \frac{\kappa}{2}(n_{th,a} + 1)(2a\rho a^\dagger - a^\dagger a \rho - \rho a^\dagger a) + \frac{\kappa}{2} n_{th,a}(2a^\dagger \rho a - aa^\dagger \rho - \rho a a^\dagger)$ and $L_b(\rho) = \frac{\gamma}{2}(n_{th,b} + 1)(2b\rho b^\dagger - b^\dagger b \rho - \rho b^\dagger b) + \frac{\gamma}{2} n_{th,b}(2b^\dagger \rho b - bb^\dagger \rho - \rho b b^\dagger)$ are the Liouvillian operators for the optical and phonon modes respectively [10]. $\gamma$ is the phonon decay rate. $n_{th,a}$ and $n_{th,b}$ are the thermal photon and phonon numbers given by $n_{th,i} = [exp\{\hbar \omega_i / k_B T\} - 1]^{-1}$, where $k_B$ is the Boltzmann constant. Thermal photon number is considered to be zero due to high frequency of optical radiation.

### III. RESULTS AND DISCUSSIONS

The steady-state value of $g^{(2)}(0)$ could be treated as a figure of merit in order to quantify single photon blockade behavior [11]. The condition $g^{(2)}(0) < 1$ corresponds to sub-Poisson feature of the cavity field, which is a nonclassical effect, referred to as photon antibunching or photon blockade, while $g^{(2)}(0) > 1$ indicates super-Poissonian statistics which is a classical effect. On the other hand, $g^{(2)}(0) \to 0$ refers to the case of full photon blockade.

For our numerical calculations we choose, in the weak single-photon optomechanical coupling regime i.e. $g < \kappa$, the following experimentally realistic parameters [16,17]: $\omega_a/2\pi = 205.6$ THz, $\lambda_a = 1.55$ μm, $\omega_b/2\pi = 9.5$ GHz, $g/2\pi = 292$ kHz, $Q_a = 10^7$, $\gamma = 0.001\kappa$. Fig. 2 depicts $g^{(2)}(0)$ as a function of the normalized cavity detuning for different values of the ratio $D$. It should be noted that, as reported in literatures, the magnitude of $\chi^{(3)}$ tensor elements for typical semiconductors are of the order of $10^{-18} -$ $10^{-19} \, m^2/V^2$ while for some nano-particle doped glasses its of the order of $10^{-16} \, m^2/V^2$ in near-infrared [18,19]. The value of $\bar{n}_r$ typically varies between 2 to 4 [20].

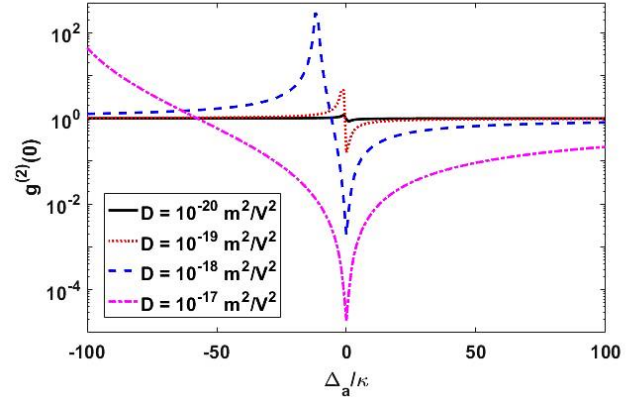

Fig. 2. Variation of zero-time-delay second-order correlation function with normalized cavity detuning for different values of $D = \bar{\chi}^{(3)}/\bar{\varepsilon}_r^2$.

The cavity mode volume is taken to be $V_{eff} = 0.01$ μm$^3$, while $\Omega = 0.01\kappa$. The figure exhibits strong photon antibunching near zero cavity detuning in the system, the maximum occurring for highest value of $D$. It should be noted that unlike bulk media, here even at low optical intensity, the nonlinear susceptibility is significant due to low mode volume, thanks to recent advances in nanofabrication techniques. The role of the cavity mode volume $V_{eff}$ on $g^{(2)}(0)$ is demonstrated in Fig. 3. Here we have assumed $D = 10^{-17} m^2/V^2$. For higher values of $V_{eff}$, $g^{(2)}(0)$ value tends to show Poissonian statistics. However, as $V_{eff}$ is lowered, the cavity-field tends to show sub-Poissonian characteristics indicating photon antibunching. These results indicates that, with judicious engineering, it is possible to obtain photon blockade in this system even if we have $g < \kappa$.

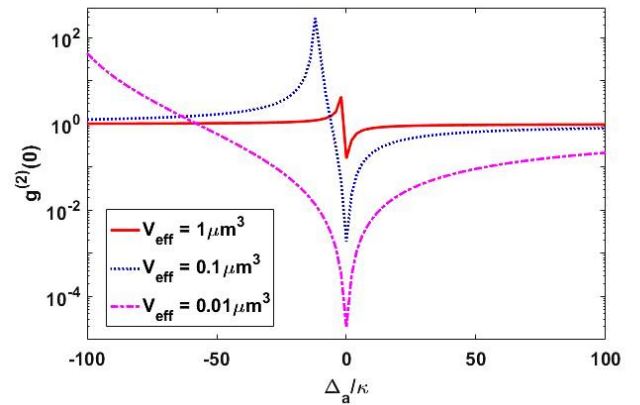

Fig. 3. Variation of $g^{(2)}(0)$ with normalized cavity detuning for different values of effective confinement volume $V_{eff}$.

The role of Kerr-nonlinear interaction, $U$, on photon antibunching is illustrated in Fig. 4. In the inset, the variation of $g^{(2)}(0)$, at $\Delta_a = 0$, with normalized $U$ is plotted. It can be seen that when $U = 0$, i.e. in absence of the nonlinear photonic crystal, $g^{(2)}(0) = 1$, referring to Poissonian photon statistics, a classical behavior. Thus, one cannot have photon antibunching, in a bare cavity, in the regime $g < \kappa$ as predicted by Rabl and others [11,12]. But with the addition of Kerr nonlinearity, it is possible to observe photon antibunching.

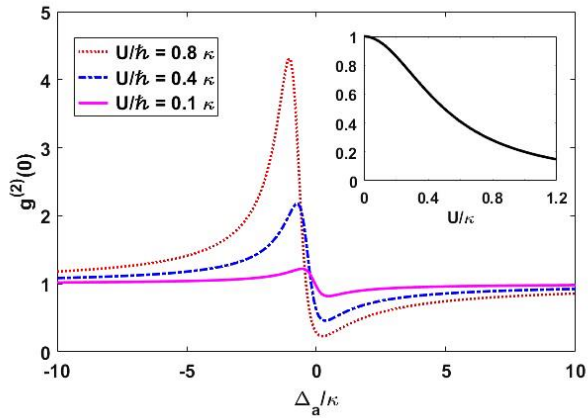

Fig. 4. $g^{(2)}(0)$ as a function of normalized cavity detuning for different values of normalized Kerr-nonlinearity strength, $U/\hbar\kappa$. In the inset, the variation of $g^{(2)}(0)$ with $U/\hbar\kappa$ is shown.

It is evident from Fig. 4 that with our proposed scheme, photon antibunching in an optomechanical cavity could be obtained in the regime where both $U/\hbar < \kappa$ and $g < \kappa$. This result may be considered significant, in view of the fact that most of the studies on photon blockade consider strong coupling regime, which is very hard to achieve in experiments. It is worthwhile to note that even though strong Kerr nonlinearity enhances antibunching effect, as could be observed from Fig. 4, useful antibunching could still be obtained in the weak Kerr nonlinearity and single-photon optomechanical-coupling regime.

## IV. Conclusion

To conclude, we have presented a scheme for realizing single-photon blockade in a weakly driven hybrid cavity optomechanical system. Photon antibunching effect is investigated, with experimentally realistic parameters, by calculating the zero-time-delay second-order correlation function, $g^{(2)}(0)$, of the cavity optical field via solving the quantum master equation numerically. It is shown that, it is possible to achieve useful photon antibunching even when the single-photon optomechanical coupling strength is smaller than the decay rate of the optical mode. The scheme proposes to relax the requirement of strong coupling for photon blockade in optomechanical systems. It is also shown that photon blockade could be generated at 1550 nm, the telecommunication wavelength, that is useful for long distance communication.